\documentclass[preprint,12pt]{elsarticle}
\usepackage{lineno,hyperref}
\usepackage{amsmath,amssymb}
\usepackage[english]{babel}
\usepackage{bm}
\usepackage{color}

\usepackage{graphicx} 
\usepackage{fancyhdr}
\usepackage{caption}

\usepackage{rotating}
\usepackage{color}

\journal{Chemical Physics Letters}

\usepackage{amsmath,amssymb}
\usepackage[english]{babel}
\usepackage{bm}
\usepackage{color}
\usepackage{indentfirst}

\newcommand{\angstrom}{\mbox{\normalfont\AA}}
 
\usepackage[normalem]{ulem}
\usepackage{color}  

\begin{document}

\begin{frontmatter}

\title{Towards reliable calculations of thermal rate constants: ring polymer molecular dynamics for the OH + HBr $\to$ Br + H$_2$O reaction}

\author[skoltechaddress]{Ivan S. Novikov\corref{correspondingauthor}}

\cortext[correspondingauthor]{Corresponding author}
\ead{ivan.novikov0590@gmail.com}

\author[skoltechaddress]{Edgar M. Makarov}

\author[yuraaddress]{Yury V. Suleimanov}

\author[skoltechaddress]{Alexander V. Shapeev}

\address[skoltechaddress]{Skolkovo Institute of Science and Technology, Skolkovo Innovation Center, Nobel St. 3, Moscow 143026, Russia}
\address[yuraaddress]{Computation-based Science and
Technology Research Center, Cyprus Institute, 20 Kavafi Street,
Nicosia 2121, Cyprus}

\begin{abstract}

We combined Moment Tensor Potential (MTP) and Ring Polymer Molecular Dynamics (RPMD) for calculating the thermal rate constants of the OH + HBr system. We used the active learning (AL) algorithm for constructing a training set during RPMD. We compared the obtained RPMD-AL-MTP rate constants with the ones previously calculated using the quasi-classical trajectories (QCT) and the POTLIB potential energy surface, and with the experimental ones. We demonstrated that the RPMD rate constants were systematically closer to the experimental rate constants than the QCT ones at 200 K, 300 K, and 500 K. 

\end{abstract}


\end{frontmatter}

\section{Introduction}

Machine-learning interatomic potentials (MLIPs) are gradually becoming a popular practical choice for molecular simulations in physics, chemistry, and materials science. One of such widely used functional forms of MLIPs is the Moment Tensor Potential (MTP) which was proposed in 2016 for single-component materials \cite{Shapeev2016-MTP} and later, in 2018, generalized to the case of multi-component materials \cite{Gubaev-2018}. A fully automated method was proposed for on-the-fly construction of potential energy surfaces (PESs) during the calculations of thermal rate constants of gas-phase chemical reactions. The method is based on combining (1) one of the most reliable full-dimensional methods for calculating thermal rate constants, Ring Polymer Molecular Dynamics (RPMD \cite{Manolopoulos2005,RPMDreview2016}), that was implemented in the RPMDrate code  \cite{RPMDrate} and successfully applied to many prototypical gas-phase chemical reactions \cite{RPMDreview2016,Suleimanov2017A,
Suleimanov2017B,Manolopoulos2011,espinosa2020vtst,suleimanov2014XH2,
rampino2016thermal,
hickson2017experimentalcd2,hickson2017oh2,bhowmick2018,menendez2019new,
nunez2018combined,del2019quantum,kumar2018low,
nunez2019experimental,hickson2021experimental}, and (2) the active learning (AL)  \cite{Podryabinkin-2017,Gubaev2019-active-mlip} of MTP (AL-MTP) algorithm implemented in the MLIP-2 code \cite{Novikov-MLIP2020}. So far, the combination of RPMD and AL-MTP methods was successfully validated during model rate calculations for two representative thermally activated chemical reactions, namely, OH + H$_2$ $\to$ H + H$_2$O \cite{Novikov-RPMD2018} and CH$_4$ + CN $\to$ CH$_3$ + HCN  \cite{Novikov-RPMD2018}, and one prototypical barrierless chemical reaction S($^1$D) + H$_2$ $\to$ HS($^1$D) + H \cite{Novikov-RPMD2019} in which AL was parameterized to the data from the previously generated analytical PESs. 

It is worth noting that, before the development of MLIPs, several methods for constructing analytical PESs were proposed and the corresponding libraries were implemented. Such PESs were mainly based on the fitting of polynomials to {\it ab initio} data. In 2002, a modified Shepard interpolation \cite{Farwig-Shepard1986} of PESs  was successfully applied to various chemical reactions. This approach is implemented in the Grow package \cite{Collins-PES2002}. In 2009, the permutationally invariant PESs in Morse-type variables of all the internuclear distances were proposed. The accuracy of these PESs were verified for such systems as CH$_5^+$, (H$_2$O)$_2$, and CH$_3$CHO \cite{Bowman-PES2009}. Since 2001, PESs created by various researchers for a variety of chemical systems have been collected in the POTLIB package as a consistent set of computer codes. The package contains PESs for a number of triatomic and polyatomic chemical systems \cite{POTLIB-2001}. In 2014, a permutationally invariant PES \cite{Bowman-PES2009} for the OH + HBr $\to$ Br + H$_2$O reaction was constructed and added to POTLIB. Quasiclassical trajectories (QCT) were used to calculate this chemical reaction rate over the temperature range of 5--500 K \cite{Oliveira-rate2014,Oliveira-PES2014}. 

Let us take a closer look at the last example. The OH + HBr $\to$ Br + H$_2$O reaction represents a significant challenge for dynamics simulations due to the quantum effects of nuclear motions during a light atom (hydrogen) transfer along a rather non-trivial reaction energy path with both an energy barrier and a pre-barrier complex formation region. This reaction is of great importance for stratospheric chemistry as it is the major process by which atomic bromine, an active ozone depletion catalyst, is regenerated from the relatively stable hydrogen bromide \cite{Yung-stratosphere1980}. This reaction is also important in combustion chemistry where it has been shown to contribute to the mechanism by which some brominated compounds act as fire retardants \cite{Clark-combustion1970}. Therefore, this system is not only interesting from the point of view of validating theoretical approaches for calculating the rate constants, but also has practical importance.

In this paper we extended the combination of RPMD and AL-MTP to the problem of direct parameterization to the electronic structure {\it ab initio} data and use this combination for predicting the OH + HBr $\to$ Br + H$_2$O reaction rate constants at different temperatures. More specifically, we calculated the RPMD rate constants using (1) the POTLIB PES, (2) the MTP fitted to the POTLIB PES energies, and, finally, (3) the MTP fitted to the electronic structure {\it ab initio} energies. We compared these rate constants with the ones previously calculated with the QCT method using the POTLIB PES \cite{Oliveira-PES2014}, and with the experimental rate constants.

\section{Methods} 

\subsection{Moment Tensor Potentials}

MTP is a local potential in the sense that the energy of a configuration $\bm x$ is partitioned into a sum of contributions of each of $N$ atoms 
\begin{equation}\label{MTP}
E^{\rm MTP} = \sum_{i=1}^N V_i = \sum_{i=1}^N \sum \limits_{\alpha} \xi_{\alpha} B_{\alpha}(\bm r_i),
\end{equation}
where $\xi_\alpha$ are the linear parameters that are found during MTP fitting and $B_{\alpha}(\bm r_i)$ are the basis functions that depend on the $i$-th atomic environment $\bm r_i$ consisting of all $j$-th atoms that are within the distance of $R_{\rm cut}$ from the $i$-th atom, i.e. $|r_{ij}| < R_{\rm cut}$.

For construction of the basis functions $B_{\alpha}$ we introduce the moment tensor descriptors \cite{Shapeev2016-MTP}
\begin{equation}\label{Moments}
M_{\mu,\nu}(\bm r_i)=\sum_{j} \underbrace {r_{ij}^{\otimes \nu}}_\text{angular part} \underbrace {\sum_{\beta} c_{\mu, z_i, z_j}^{(\beta)} \varphi^{(\beta)}(|r_{ij}|)}_\text{radial part}.
\end{equation}
Here the symbol ``$\otimes$'' denotes the outer product of the relative atomic positions $r_{ij}$, $z_i$ is the type of the $i$-th atom, and $z_j$ is the type of the $j$-th neighboring atom. The first part in \eqref{Moments}, $r_{ij}^{\otimes \nu}$, is the angular part, it describes many-body interactions. The second part in \eqref{Moments} is the $\mu$-th radial part which determines two-body interactions. We note that the radial part is the expansion over the polynomials $\varphi^{(\beta)}$ which tends to zero when the distance between atoms is close to a cut-off radius $R_{\rm cut}$. The weights $c_{\mu, z_i, z_j}^{(\beta)}$ of this expansion is another set of MTP parameters to be fitted. 

We then contract these descriptors $M_{\mu,\nu}$ to a scalar and obtain rotationally-invariant basis functions $B_{\alpha}$. To define a particular functional form of MTP, we introduce the so-called level of the moment tensor descriptors ${\rm lev} M_{\mu,\nu} = 2 + 4 \mu + \nu$, choose the maximum level, ${\rm lev}_{\rm max}$, and include in \eqref{MTP} only the basis functions with ${\rm lev} B_{\alpha} \leq {\rm lev}_{\rm max}$ (see \cite{Novikov-MLIP2020} for details). We denote the parameters of MTP to be fitted by ${\bm {\theta}}$ and the MTP energy of a configuration $\bm x$ by $E^{\rm MTP} = E({\bm {\theta}}; \bm x)$.

\subsection{PES models}

In this paper we use two PES models. The first PES model (and, also, reference) model is the PES proposed in \cite{Bowman-PES2009}. This model was fitted to the electronic structure {\it ab initio} data (we describe the level of theory in the next paragraph) in \cite{Oliveira-rate2014,Oliveira-PES2014} and used to predict the rate constants of the OH + HBr $\to$ Br + H$_2$O reaction with the quasi-classical trajectories (QCT) method. The PES has the form:
\begin{equation} \label{POTLIB_PES}
\displaystyle
V(y_1,\ldots,y_6)=\sum \limits_{n_1,\ldots,n_6}C_{n_1,\ldots,n_6}y_1^{n_1}y_6^{n_6}[y_2^{n_2}y_3^{n_3}y_4^{n_4}y_5^{n_5}+ y_2^{n_4}y_3^{n_5}y_4^{n_2}y_5^{n_3}],
\end{equation}
where $y_i = e^{-r_i/2a_0}$, $r_i$ is an internuclear distance, namely: $r_1 = r_{\rm HH'}$, $r_2 = r_{\rm HO}$, $r_3 = r_{\rm HBr}$, $r_4 = r_{\rm H'O}$, $r_5 = r_{\rm H'Br}$, and $r_6 = r_{\rm OBr}$, i.e., the PES is invariant with respect to the hydrogen atom permutations. The order of the polynomials in \eqref{POTLIB_PES} is $n_1+\ldots+n_6 \leq 6$, thus, we have 502 parameters. This PES is published in the open-source POTLIB library. We refer to this PES as POTLIB. We use it in our first test, namely, for comparison of the rate constants obtained with the MTP fitted to the POTLIB PES with the ones obtained with POTLIB \cite{Oliveira-rate2014,Oliveira-PES2014} at different temperatures. 

The second PES model is the electronic structure one which is based on the present calculations based on the spin-unlimited method of explicitly correlated coupled clusters with allowance for double excitations and partial allowance for triple excitations (UCCSD(T)-F12a). The electronic structure {\it ab initio} calculations were carried out using the MOLPRO package. The restricted Hartree-Fock method for open shells (ROHF) was used to construct the reference functions. The calculations were performed using the cc-pVDZ-F12 orbital basis set and the following subsets: cc-pVDZ(-PP)-F12/OptRI for identity resolution, cc-pVTZ/JKFIT (for H and O), and QZVPP/JKFIT (for Br) when fitting the density of the exchange operator and the Fock operator, as well as the sets aug-cc-pVTZ/MP2FIT (for H and O) and cc-pVTZ-PP-F12/MP2FIT (for Br) for the remaining two-electron integrals. Instead of explicitly taking into account the electrons of the lowest energy levels of bromine, the effective relativistic ten-electron potential (ECP10MDF) was used. The specific choice of the method and the orbital and auxiliary basis sets were the same as in \cite{Oliveira-PES2014} and correspond to a high level of the electronic structure theory. We note that the POTLIB PES \cite{Oliveira-PES2014} used here and the MTP PES were fitted at the same level of theory.

\subsection{RPMD-AL-MTP algorithm with high level of the electronic structure theory}

\begin{figure}[h!]
\centering
\includegraphics[width=5.3in, height=3.0in, keepaspectratio=false]{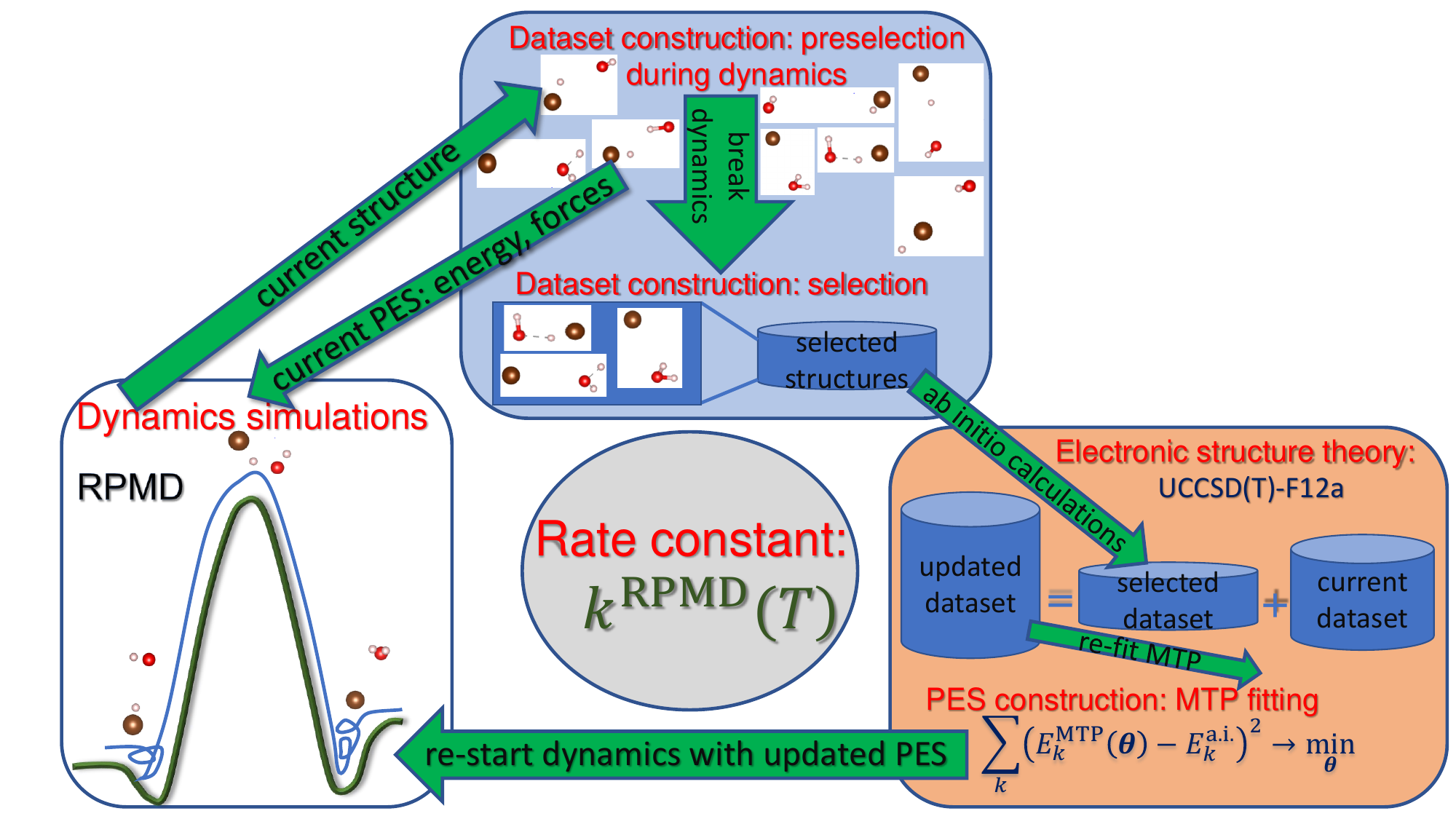}
\caption{\label{Fig:TOC} Scheme of the methodology for reliable estimation of thermal rate constants based on combining the best practices for dynamic simulations (RPMD), PES construction (AL-MTP), and the electronic structure calculations (UCCSD(T)-F12a). The RPMD method gives structures, we calculate their energies and forces with the current MTP and preselect some of them during dynamics. At some point we stop RPMD simulation according to the AL method, select some of the preselected structures, and conduct {\it ab initio} calculations for them. Finally, we update the training set (dataset) by adding the selected dataset to the current dataset, re-fit the current MTP, and re-start RPMD with the updated MTP. We continue this cycle until we obtain the final thermal rate constants.}
\end{figure}

Assume there are $K$ starting configurations in the training set. Each configuration ${\bm x^{(k)}}$ in the training set has {\it ab initio} energy precalculated. We denote {\it ab initio} energies by $E^{\rm a.i.}$. For finding the parameters $\bm \theta$, i.e., MTP fitting, we minimize the loss function
\begin{equation} \label{Fitting}
\begin{array}{c}
\displaystyle
L({\bm {\theta}}) = \sum \limits_k \left(E({\bm {\theta}}; \bm x^{(k)}) - E^{\rm a.i.}(\bm x^{(k)}) \right)^2 \to \operatorname{min}.
\end{array}
\end{equation} 
Starting from $K$ initial configurations we create the resulting training set automatically using the AL \cite{Podryabinkin-2017,Gubaev-2018} algorithm during the RPMD \cite{Manolopoulos2005} simulation. The AL algorithm allows us to select configurations for adding to the training set on which MTP will be trained and the RPMD method allows us calculating rate constants with high accuracy. The combination of MTP active training and RPMD methods (RPMD-AL-MTP algorithm) has already been successfully applied to the thermally activated OH + H$_2$ $\to$ H + H$_2$O \cite{Novikov-RPMD2018} and CH$_4$ + CN $\to$ CH$_3$ + HCN \cite{Novikov-RPMD2018} chemical reactions, and the barrierless S($^1$D) + H$_2$ $\to$ HS($^1$D) + H \cite{Novikov-RPMD2019} chemical reaction. However, in the mentioned studies MTP was fitted to the energies and forces calculated with the reference model PESs rather than an electronic structure {\it ab initio} model. Here we go beyond the previous model calculations and also fit MTP to the data calculated with the high level of the electronic structure theory (UCCSD(T)-F12a). The scheme of our methodology is shown in Fig. \ref{Fig:TOC}. The algorithm consists of the following steps:

\begin{itemize}

\item[\bf Step 0.] Initialize the training set, fit the initial MTP.

\item[\bf Step 1.] Run the RPMD simulation with the current MTP, preselect configurations that are geometrically different from the ones in the current training set during the simulations. We note that we do not conduct the electronic structure calculations for the preselected (or, extrapolative) configurations. Stop RPMD if we meet a configuration that is geometrically very different from those in the current training set.

\item[\bf Step 2.] Select some of the preselected configurations. Typically these configurations geometrically differ from each other and maximally differ from the ones in the current training set. Such a technique allows us constructing the training set with many various configurations, i.e., we avoid adding similar configurations to the training set.

\item[\bf Step 3.] Conduct the electronic structure {\it ab initio} calculations for the selected configurations, update the current training set (i.e., add the selected configurations to the training set).

\item[\bf Step 4.] Re-train the current MTP, obtain the updated MTP.

\item[\bf Step 5.] Repeat {\bf Steps 1-4} with the updated MTP in every cycle shown in Fig. \ref{Fig:TOC} until no configuration will be added to the training set. Obtain the resulting training set and MTP.

\item[\bf Step 6.] Calculate the thermal rate constant. 

\end{itemize}

\noindent Each cycle of {\bf Steps 1-4} in the algorithm extends the training set and improves the MTP in a sense that it becomes more reliable for accurate predicting the thermal rate constant. It also allows us creating a PES ``on-the-fly'' during RPMD simulations and minimizing the number of the electronic structure {\it ab initio} calculations needed to construct a reliable PES.

\subsection{Computational details}

We use the RPMDrate code \cite{RPMDrate} for calculating the chemical reaction OH + HBr $\to$ Br + H$_2$O rate constants and the MLIP-2 code \cite{Novikov-MLIP2020} for the active training of MTP. POTLIB is available online \cite{POTLIB_PES}. We use the MOLPRO package for the electronic structure calculations \cite{Werner-MOLPRO2012}.

In order to compare with the experimental rate coefficients, it is necessary to take OH spin-orbit coupling effects into account, which are not included in RPMD calculations. The $^2 \Pi_{1/2}$ excited state of OH, which is $\Delta = 140 ~{\rm cm^{-1}}$ above the $^2 \Pi_{3/2}$ ground state can be included by considering that the ratio of transition state to reactant electronic partition functions adds a factor $F(T) = 1 + e^{-\Delta/k_B T}$, as in the previous theoretical studies. The resulting spin-orbit corrected rate constant is $k_{\rm RPMD}^{\rm corr.}(T) = F^{-1}(T) \times k_{\rm RPMD}(T)$. We also compare the RPMD rate constants with the QCT rate constants taken from \cite{Oliveira-PES2014}. 

We calculate the RPMD rate constants at $T = 200$, 300, and 500 K and consider 192, 128, and 96 ring polymer beads, respectively. In order to obtain the potential of mean force (PMF) $W(\xi)$ along the reaction coordinate $\xi$ we divide the interval of the coordinate $-0.05 \leq \xi \leq 1.05$ into 111 windows of width 0.01. We determine the dynamical correction to the rate constant, namely, the recrossing factor, at the ``plateau'' time observed around 0.3 ps for $T = 200$, 300, and 500 K. 

We run the calculations with the RPMD method using two PESs: POTLIB and MTP fitted to the electronic structure {\it ab initio} calculations with the MOLPRO package (we denote it by MTP$\_$MOLPRO). To additionally test MTP we fit it to POTLIB (we denote this PES by MTP$\_$POTLIB) during RPMD and compare the resulting rate constants obtained with POTLIB and MTP$\_$MOLPRO PESs. Each MTP contains 527 parameters to optimize (${\rm lev}_{\rm max} = 16$) and the cut-off radius of 6 $\angstrom$. Average calculation time on a single central processing unit (CPU) core is about 0.6 microseconds per atom for POTLIB and 17 microseconds per atom for MTP.

In addition, we calculate classical rate constants, i.e. with one ring polymer bead using the POTLIB PES and MTP fitted to MOLPRO at $T = 200$, 300, and 500 K. With these calculations, we demonstrate the importance of incorporating quantum mechanical effects using the RPMD method to accurately predict rate constants and their trend with decreasing temperature.

\section{Results and discussion}

\subsection{Potential of mean force}

\begin{figure}[h!]
\centering
\includegraphics[width=5.5in, height=6.5in, keepaspectratio=false]{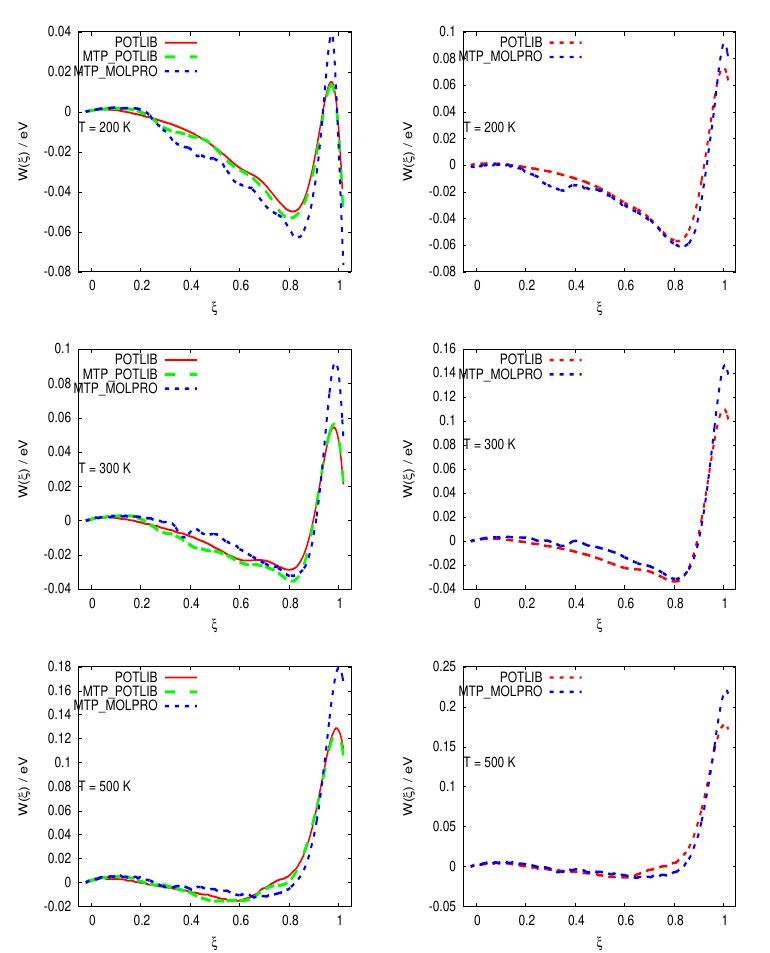}
\caption{\label{Fig:PMF} Potentials of mean force obtained with RPMD (left) and classical MD (right) using different PESs at $T = 200$, 300, and 500 K.}
\end{figure}

The PMF profiles obtained with the RPMD method using POTLIB, MTP$\_$POTLIB, and MTP$\_$MOLPRO PESs at $T = 200$, 300, and 500 K are shown in Fig. \ref{Fig:PMF} on the left. We see that the free energy  has a non-trivial profile along the reaction path which includes both pre-reaction complex formation well and thermally activated free energy barrier. The interplay of these two factors depends on temperature -- while at 500 K the profile is close to a standard one for typical thermally activated reactions, at 200 K the complex-formation region plays an important role. 

As it was expected, the profiles obtained with the POTLIB and MTP$\_$POTLIB PESs are close to each other at $T = 200$, 300, and 500 K temperatures. The profiles calculated with MTP$\_$MOLPRO turned out to be unexpected: the difference (or, the free energy barrier) between the minimum free energy and the one at the saddle point is much larger for the MTP$\_$MOLPRO profile than the similar value for the POTLIB and MTP$\_$POTLIB profiles at $T = 200$, 300, and 500 K. 

The classical PMF profiles obtained using POTLIB and MTP$\_$MOLPRO PESs at $T = 200$, 300, and 500 K are given in Fig. \ref{Fig:PMF} on the right. From the classical PMF profiles we conclude that the reaction barriers predicted with classical MD are higher than the ones calculated with RPMD. This fact leads to the significant difference between the classical and RPMD rate constants. Thus, Fig. \ref{Fig:PMF} demonstrates the importance of accounting for quantum mechanical effects in the OH + HBr $\to$ Br + H$_2$O reaction.

\subsection{Rescrossing factor}

\begin{figure}[h!]
\centering
\includegraphics[width=5.5in, height=6.5in, keepaspectratio=false]{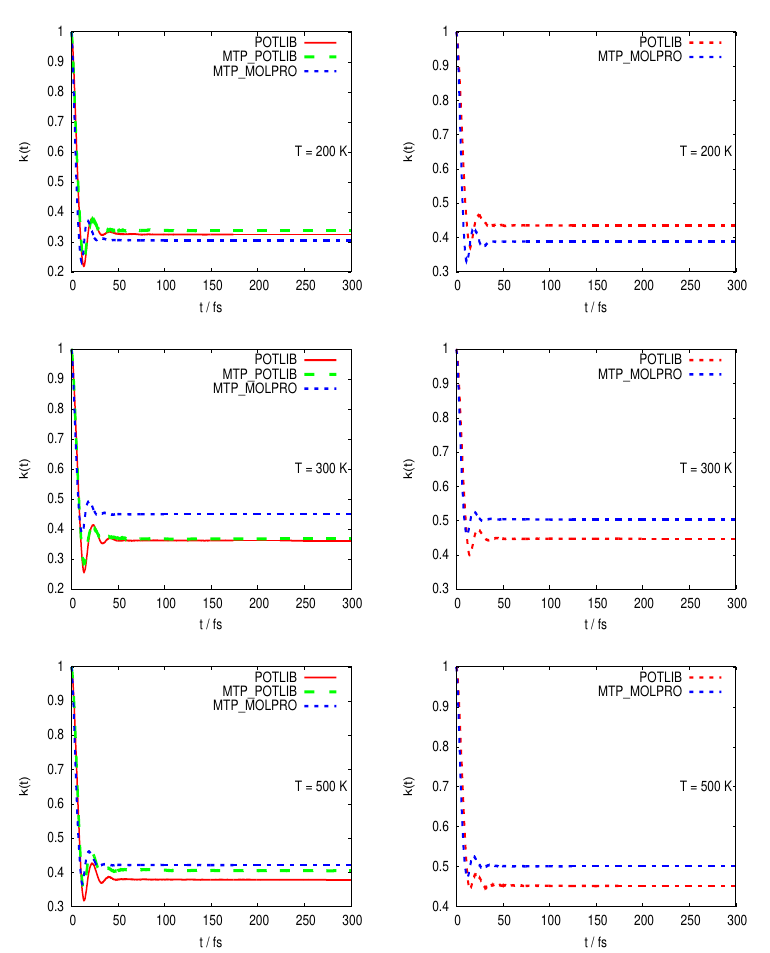}
\caption{\label{Fig:TC} Recrossing factors obtained with RPMD (left) and clasical MD (right) using different PESs at $T = 200$, 300, and 500 K.}
\end{figure}

The profiles of the recrossing factor obtained with the RPMD method using different PESs at $T = 200$, 300, and 500 K are demonstrated in Fig. \ref{Fig:TC} on the left. As in the case of PMF profiles, the recrossing factors obtained with the POTLIB and MTP$\_$POTLIB PESs are close to each other. The MTP$\_$MOLPRO recrossing factor at $T = 300$ K is slightly larger than the ones obtained with the PESs on the basis of POTLIB, however, the relative difference is not as large as for the free energy. The ``plateau'' time is equal to 300 fs for all the temperatures. Due to the strong quantum nature of hydrogen transfer in the title reaction it would be logical to extend the present calculations to lower temperatures such as 50 K and 100 K. However, during the calculations at $T = 100$ K, we encountered difficulties in calculating the ring polymer recrossing factor. In particular, the ``plateau'' time exceeded 100 ps, which made calculations of chemical reaction rate constants at low temperatures resource-intensive and these results are not presented in this work.

The profiles of the recrossing factor obtained with calssical MD using the POTLIB and MTP$\_$MOLPRO PESs are shown in Fig. \ref{Fig:TC} on the right. Here we observe that the classical profiles calculated with POTLIB and MTP$\_$MOLPRO are close to each other at all the temperatures considered, the ``plateau'' time is similar for all the recrossing factors.
 
\subsection{Comparison of the training sets and the fitting errors}

Here we provide the information on the resulting training sets created for fitting the MTP$\_$MOLPRO PESs obtained during RPMD and classical MD and the fitting energy RMSEs. The training sets for the MTP$\_$MOLPRO PESs fitting were automatically constructed. First, we uniformly initialized the training set with 1000 configurations along the reaction coordinate on the interval $-0.05 \leq \xi \leq 1.05$. For sampling the coordinates of the configurations in the initial training set we used the POTLIB PES during the classical MD simulations at $T = 300$ K and for single-point {\it ab initio} calculations of the energies of these configurations we used the MOLPRO package. We fitted the initial MTP on this training set. Next, starting from this initial training set and the MTP fitted on it we ran the RPMD-AL-MTP algorithm described above and automatically created the resulting training sets and the MTPs at different temperatures. We note, that our idea is to create any PES ``on-the-fly'' at each particular temperature, i.e., we do not use the PES and the training set created for one temperature as an initial guess for the other temperatures. In principle, it is possible, but here we test our ``fully automated methodology'' starting only from 1000 initial configurations at any temperature. The sizes of the training sets and the corresponding fitting energy RMSEs for the MTP\_MOLPRO PESs obtained during RPMD and classical MD are given in Table \ref{Tab:FittingErrors}. From the table we conclude that we need more configurations at high temperatures than at low temperatures for MTP fitting during classical MD as apposed to the fitting during RPMD. This is due to the fact that we neglect quantum-mechanical effects in classical MD and the deviations of atoms from the equilibrium positions become greater. We also see that the fitting errors for the MTPs trained during classical MD are smaller than the ones for the MTPs trained during RPMD. This is also a consequence of the fact that classical dynamics does not take quantum mechanical effects into account, therefore, the variety of configurations in the ``classical'' traing sets is less than in the ones obtained during RPMD and, thus, the training errors are smaller. Overall, the fitting errors for all the potentials were smaller than 43 meV, i.e., 1 kcal/mol corresponding to the ``reasonable'' chemical accuracy.

\begin{table}[h!]
\begin{center}
\caption{\label{Tab:FittingErrors} Training set sizes and fitting energy RMSEs for the MTP\_MOLPRO PESs obtained during RPMD and classical MD at different temperatures.} 
\begin{tabular}{c|c|c|c} \hline \hline
 Method & Temperature & Training set size & energy RMSE, meV 
 \\ \hline
 & $T = 200$ K & 8078 & 22 
  \\ 
RPMD & $T = 300$ K & 7535 & 20
  \\ 
 & $T = 500$ K & 7671 & 25
  \\ \hline
 & $T = 200$ K & 6502 & 15 
  \\ 
classical MD & $T = 300$ K & 7376 & 14
  \\ 
 & $T = 500$ K & 7413 & 14
  \\ 
\hline
\hline
\end{tabular}
\end{center}
\end{table}

\subsection{Comparison of the minima and saddle points obtained with different models}

\begin{figure}[h!]
\centering
\includegraphics[width=5.0in, height=6.5in, keepaspectratio=false]{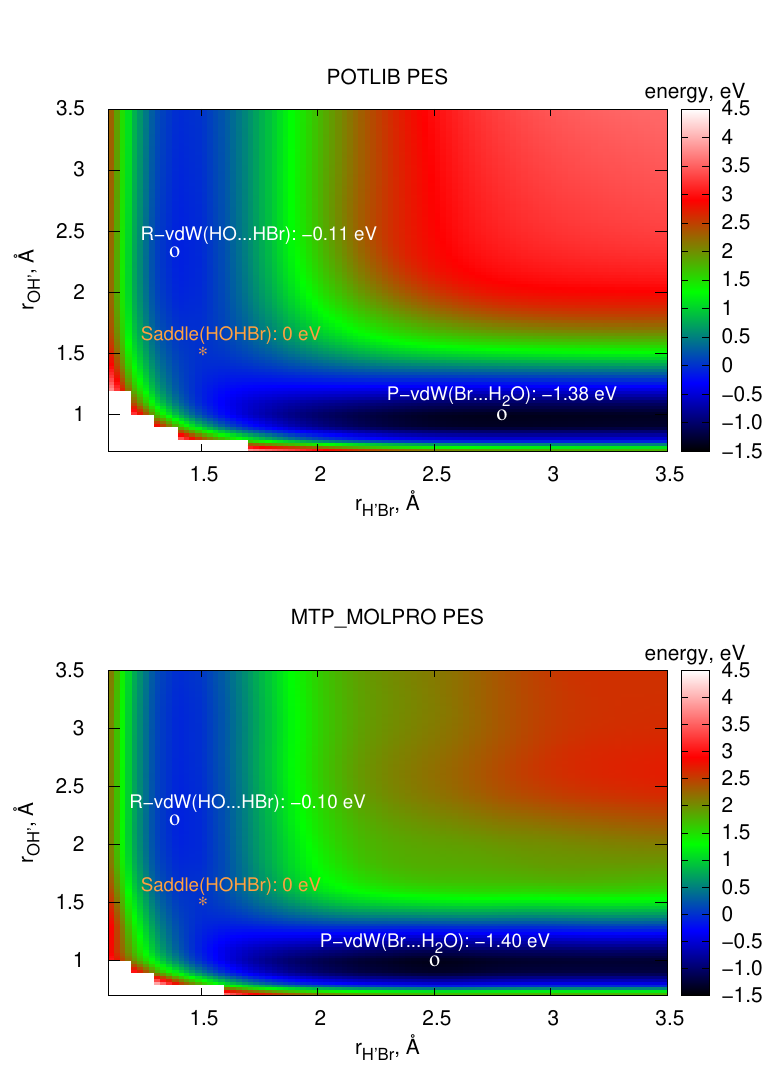}
\caption{\label{Fig:Contours} Equipotential contour plots obtained with the POTLIB PES and with the MTP$\_$MOLPRO PES at $T = 300$ K. The POTLIB PES reproduces the contour in Fig. 2 from \cite{Oliveira-PES2014}. Both the POTLIB PES and the MTP$\_$MOLPRO PES were obtained with the same level of electronic structure theory.}
\end{figure}

\begin{figure}[h!]
\centering
\includegraphics[width=5.8in, height=1.1in, keepaspectratio=false]{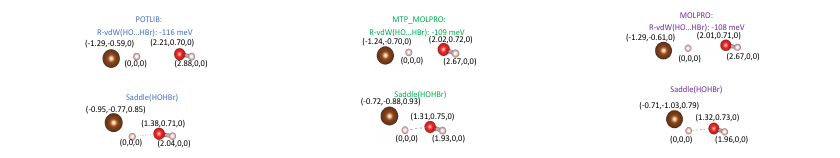}
\caption{\label{Fig:MinSaddle} Configurations in R-vdW(HO$\cdots$HBr) minima and saddle(HOHBr) points obtained with the POTLIB PES, the MTP$\_$MOLPRO PES at $T = 300$ K, and MOLPRO. The coordinates of atoms are given in $\angstrom$. The energies in the minima are given with respect to corresponding reference energies in the saddle points that are considered to be 0 eV. The POTLIB van der Waals minimum is geometrically different from the ones obtained with the MTP$\_$MOLPRO PES and MOLPRO, whereas the saddle points are close to each other.}
\end{figure}

The relaxed equipotential contour plots obtained with the POTLIB PES (constructed during QCT) and the MTP$\_$MOLPRO PES (fitted during RPMD) at $T = 300$ K are shown in Fig. \ref{Fig:Contours}. From the figure we conclude that the PESs give rather different contour plots. The R-vdW(HO$\cdots$HBr) and P-vdW(Br$\cdots$H$_2$O) minima calculated with the POTLIB and MTP$\_$MOLPRO PESs are located in different zones, however the saddle(HOHBr) points are close to each other. We demonstrate the same fact in Fig. \ref{Fig:MinSaddle} and Table \ref{Tab:MinSaddleGeometry} where the exact R-vdW(HO$\cdots$HBr) minima and the saddle(HOHBr) points calculated with the PESs and the MOLPRO code are shown. Here we also see that the saddle(HOHBr) points obtained with the POTLIB PES, the MTP$\_$MOLPRO PES, and with MOLPRO are close to each other. However, the R-vdW(HO$\cdots$HBr) minimum obtained with the POTLIB PES differs from the ones obtained with the MTP$\_$MOLPRO PES and MOLPRO. Nevertheless, in Fig. \ref{Fig:Contours}, Fig. \ref{Fig:MinSaddle}, and Table \ref{Tab:MinSaddleGeometry} we demonstrate that both POTLIB and MTP$\_$MOLPRO do not produce obviously ``spurious'' results in critically important points and regions like the reactants zone, the zone around the transition state, or the products zone, and, thus, we show that the POTLIB and MTP$\_$MOLPRO PESs are applicable to investigating the OH+HBr system. There is a fairly large difference between the POTLIB and the MTP$\_$MOLPRO contours in the region where $r_{\rm OH'} > 2 ~\angstrom$ and $r_{\rm H'Br} > 2 ~\angstrom$: the MTP$\_$MOLPRO PES gives smaller energies than the POTLIB PES. We note that all the configurations on the contour plots were created manually, not during RPMD. Thus, some of the configurations like the ones with $r_{\rm OH'} > 2 ~\angstrom$ and $r_{\rm H'Br} > 2 ~\angstrom$ were not included to the training set automatically obtained with the AL algorithm during RPMD and, therefore, MTP extrapolates for the configurations in this region. However, since these configurations did not occur during the RPMD simulation, they do not affect the accuracy of the prediction of chemical reaction rates and, therefore, it is not necessary to have accurate PESs in this region. Both models demonstrate that the title reaction features a submerged barrier with a negative energy with respect to the OH+HBr asymptote. We emphasize that the contour plots for the other two MTP$\_$MOLPRO PESs at $T = 200$ K and $T = 500$ K are close to the one at $T = 300$ K.  Therefore, we do not demonstrate these contour plots here.

\begin{table}[h!]
\begin{center}
\footnotesize
\caption{\label{Tab:MinSaddleGeometry} Geometries of R-vdW(HO$\cdots$HBr) minima and saddle(HOHBr) points obtained with the POTLIB PES, the MTP$\_$MOLPRO PES at $T = 300$ K, and MOLPRO.} 
\begin{tabular}{c|c|c|c|c|c|c|c} \hline \hline
Point & Model & $r_{\rm HO}$ & $r_{\rm OH'}$ & $r_{\rm H'Br}$ & $\theta_{\rm HOH'}$ & $\theta_{\rm OH'Br}$ & $\phi_{\rm HOH'Br}$
 \\ \hline
 & POTLIB & 0.97 & 2.32 & 1.42 & 116.5 & 173.2 & 0 
  \\ 
R-vdW(HO$\cdots$HBr) & MTP$\_$MOLPRO & 0.97 & 2.14 & 1.43 & 112.8 & 170.0 & 0 
  \\ 
 & MOLPRO & 0.97 & 2.13 & 1.42 & 113.6 & 174.2 & 0  
 \\ \hline
 & POTLIB & 0.97 & 1.55 & 1.49 & 105.7 & 143.7 & 73.7 
  \\ 
Saddle (HOHBr) & MTP$\_$MOLPRO & 0.97 & 1.52 & 1.47 & 99.6 & 136.2 & 66.0 
  \\ 
 & MOLPRO & 0.97 & 1.51 & 1.48 & 101.9 & 139.5 & 55.1 
  \\ 
\hline
\hline
\end{tabular}
\end{center}
\end{table}

In addition to the comparison of minima and saddle points, we provide an information on the saddle point obtained with MOLPRO. The PES associated with the studied chemical reaction exhibits a saddle point characterized by a classical energy minimum of $-0.34$ kcal/mol$^{-1}$ relative to the reactants. This energy value aligns precisely with the results presented in \cite{Oliveira-PES2014} for the VDZ-F12 basis set used in the present work. Upon consideration of the zero-point vibrational energies, the barrier height is established at $0.28$ kcal/mol$^{-1}$.
The vibrational frequencies of the reactants asymptote are 3737 cm$^{-1}$ for the OH and 2651 cm$^{-1}$ for HBr, which are the same as in \cite{Oliveira-PES2014}. At the saddle point, the vibrational modes have five real and one imaginary frequencies: 3719, 1764, 694, 400, 244, and 632i cm$^{-1}$. These frequency values also agree with the saddle point frequencies 3719, 1758, 689, 376, 221, and 654i cm$^{-1}$ reported in \cite{Oliveira-PES2014}.

\subsection{Theoretical and experimental thermal rate constants}

The theoretical and experimental thermal rate constants of the OH + HBr $\to$ Br + H$_2$O reaction are given in Table \ref{Tab:Rate}. The theoretical rate constants were calculated with the RPMD method using the POTLIB PES, the MTP$\_$POTLIB PES, and the MTP$\_$MOLPRO PES, and with classical MD using the POTLIB PES and the MTP$\_$MOLPRO PES. We denote them by $k_{\rm RPMD}$ and $k_{\rm cl}$, respectively. We compare the calculated $k_{\rm RPMD}$ and $k_{\rm cl}$ constants with the theoretical $k_{\rm QCT}$ constants obtained with the QCT method using the POTLIB PES and with the experimental $k_{\rm exp.}$ constants. The $k_{\rm QCT}$ and $k_{\rm exp.}$ constants were taken from \cite{Oliveira-PES2014,takacs1973reactions,jourdain1981epr,ravishankara1985oh,
ravishankara1979absolute,sims1994ultra,atkinson1997low,
jaramillo2001temperature,jaramillo2002consensus,mullen2005temperature}.

\begin{table}[h!]
\begin{center}
\caption{\label{Tab:Rate} Comparison of the thermal rate constants, obtained theoretically (with the RPMD and QCT \cite{Oliveira-PES2014} methods) and experimentally \cite{takacs1973reactions,jourdain1981epr,ravishankara1985oh,
ravishankara1979absolute,sims1994ultra,atkinson1997low,
jaramillo2001temperature,jaramillo2002consensus,mullen2005temperature}. The values of the constants are given in $10^{-11}$ cm$^3 \cdot$ (molecule)$^{-1}$ s$^{-1}$.}
\begin{tabular}{c|c|c|c} \hline \hline
$T$, K & 200 & 300 & 500 \\ \hline
$k_{\rm QCT}$ POTLIB & 2.50 $\pm$ 0.50 & 1.75 $\pm$ 0.25 & 1.40 $\pm$ 0.20 \\
$k_{\rm cl}$ POTLIB & 0.38 $\pm$ 0.08 & 0.46 $\pm$ 0.09 & 0.70 $\pm$ 0.14 \\
$k_{\rm RPMD}$ POTLIB & 8.30 $\pm$ 1.65 & 3.20 $\pm$ 0.65 & 1.70 $\pm$ 0.35 \\
$k_{\rm RPMD}$ MTP$\_$POTLIB & 9.60 $\pm$ 1.92 & 3.05 $\pm$ 0.61 & 2.06 $\pm$ 0.41 \\
$k_{\rm cl}$ MTP$\_$MOLPRO & 0.11 $\pm$ 0.02 & 0.13 $\pm$ 0.02 & 0.28 $\pm$ 0.05 \\
$k_{\rm RPMD}$ MTP$\_$MOLPRO & 2.00 $\pm$ 0.40 & 0.96 $\pm$ 0.19 & 0.63 $\pm$ 0.13 \\
$k_{\rm exp}$ & 1.35 $\pm$ 0.65 & 1.00 $\pm$ 0.20 & 1.00 $\pm$ 0.20 \\
\hline
\hline
\end{tabular}
\end{center}
\end{table}
Several important conclusions can be drawn from Table \ref{Tab:Rate}. Firstly, the $k_{\rm QCT}$ and $k_{\rm RPMD}$ rate constants calculated with the POTLIB PES are of the same order of magnitude, but there are differences in the absolute values. The rate constants are almost the same at $T = 500$ K, the difference between them is at around 20\% which corresponds to the accuracy of the RPMDrate methodology \cite{RPMDreview2016,RPMDrate}. The $k_{\rm RPMD}$ rate constant is greater than the $k_{\rm QCT}$ rate constant by factors of 1.8 and 3.3 at $T = 300$ K and $T = 500$ K, respectively. The difference between the QCT and RPMD rate constants at low temperatures is very-well known \cite{RPMDreview2016}. It originates from the difference in how these methods treat quantum-mechanical effects of nuclear motions. With RPMD they are systematically more accurate and reliable. Also, the trend of rate constant decreasing with temperature increasing was reproduced by the RPMD method. 

Secondly, we compared the $k_{\rm RPMD}$ thermal rate constants obtained with the POTLIB and MTP$\_$POTLIB PESs calculated using RPMD-AL-MTP algorithm. The rate constants at $T = 200$ K, $T = 300$ K, and $T = 500$ K are in a good agreement to each other, the accuracy of their calculation is close to the accuracy of the RPMD method. As for the POTLIB potential, the trend of rate constant decreasing with temperature increasing was reproduced by the RPMD-AL-MTP method.

Finally, we compared the theoretical and experimental thermal rate constants for the title reaction. The closest to the experimental rate constants were the ones obtained using the MTP$\_$MOLPRO: the agreement between MTP$\_$MOLPRO and the experimental constant was almost perfect at $T = $ 300 K, while the constants obtained at $T = 200$ K and $T = 500$ K with the MTP$\_$MOLPRO is closer to the experimental constants than all the other theoretical constants. We note that all the data in \cite{Oliveira-PES2014} are published up to a temperature of 450 K. However, the rate constants change insignificantly at the temperatures higher than 400 K and the comparison at $T = 500$ K makes sense. 

As it was mentioned above, we do not provide the results for the temperatures lower than 200 K due to the fact that we faced difficulties when calculating the ring polymer recrossing factor at 100 K. In this regard, the development of more efficient computational RPMDrate procedure at low temperatures for calculating the rate constants of chemical reactions in which the formation of complexes takes place is highly desirable in future. However, the present results for $T = 200-500$ K and their comparison with the previous theoretical estimates demonstrate that the proposed RPMD-AL-MTP combination is a reliable and consistent approach for calculating thermal rate constants.

\section{Conclusion}
In summary, in this paper we calculated thermal rate constants of the thermally activated OH + HBr $\to$ Br + H$_2$O chemical reaction which is of great importance for stratospheric chemistry. We used the combination of Active Learning of Moment Tensor Potential (AL-MTP) fitted to the data obtained with the spin-unlimited method of explicitly correlated coupled clusters with allowance for double excitations and partial allowance for triple excitations (UCCSD(T)-F12a) and Ring Polymer Molecular Dynamics (RPMD). The rate constants obtained with the RPMD-AL-MTP method at $T = 200$, 300, and 500 K were compared with the ones obtained previously with the method of quasi-classical trajectories (QCT) using the POTLIB potential energy surface (PES) \cite{Oliveira-PES2014} and with the experimental ones published earlier. Both the MTP and POTLIB potentials were parameterized to the data obtained at the same high level of the electronic structure theory. The rate constants calculated with the MTP fitted to the electronic structure {\it ab initio} data are in better agreement with the experimental ones compared to the previously calculated theoretical rate constants, including the QCT ones, at all three temperatures (200, 300, and 500 K). Thus, it was demonstrated that the proposed methodology based on combining the electronic structure calculations (UCCSD(T)-F12a), PES construction (AL-MTP) and dynamic simulations (RPMD) is an important step towards reliable theoretical estimation of thermal rate constants. Despite the good performance of our method at temperatures above 200 K, the RPMDrate computational procedure was found to be non-optimal at temperatures lower than 200 K for the OH + HBr $\to$ Br + H$_2$O chemical reaction. This limitation necessitates for the development of more efficient RPMDrate procedure at low temperatures for calculating the rate constants of chemical reactions in which a complex formation plays an important role.

\section*{CRediT authorship contribution statement} 

\textbf{Ivan S. Novikov}: Investigation, Visualization, Data Curation, Validation, Software, Writing - Original draft. \textbf{Edgar M. Makarov}: Investigation, Data Curation. \textbf{Yury V. Suleimanov}: Conceptualization, Methodology, Software, Writing - Review \& Editing. \textbf{Alexander V. Shapeev}: Conceptualization, Methodology, Resources, Writing - Review \& Editing, Project administration, Funding acquisition, Supervision. 
 
\section*{Declaration of Competing Interest}

None.

\section*{Acknowledgements}
This work was supported by the Russian Foundation for Basic Research (grant  number  20-03-00833). 


\end{document}